\documentclass[runningheads]{svmult}

\usepackage{makeidx}   
\usepackage{graphicx}  
\usepackage{subeqnar}  
\usepackage{multicol}  
\usepackage{physprbb}  
\makeindex             

\begin{document}
\title*{The Origin of the Correlation between the Spin Parameter and the Baryon Fraction of Galactic Disks}
\toctitle{Disk Spin Parameterand Baryon Fraction}
%
%
\titlerunning{Angular Momentum Problem}
%
\author{Andreas Burkert\inst{1}
\and Frank C. van den Bosch\inst{2}
\and Rob A. Swaters\inst{3}}
\authorrunning{Burkert et al.}
%
%
\institute{$^1$Max-Planck-Institut f\"ur Astronomie, Heidelberg, Germany \\
$^2$Max-Planck-Institut f\"ur Astrophysik, Garching, Germany \\
$^3$Carnegie Institution of Washington, Washington DC 20015, USA}

\maketitle              

\begin{abstract}
The puzzling correlation between the spin parameter $\lambda$ of galactic disks and the disk-to-halo mass
fraction $f_{disk}$ is investigated.
We show that such a correlation arises naturally from uncertainties in determining the
virial masses of dark matter halos. This result leads to the conclusion that the halo properties derived from
fits to observed rotation curves are still very uncertain which might explain part of the disagreements between
cosmological models and observations. We analyse $\lambda$ and $f_{disk}$ as function of the
adopted halo virial mass. Reasonable halo concentrations require
$f_{disk} \approx 0.01-0.07$ which is significantly smaller than the universal baryon fraction.
Most of the available gas either never settled into the galactic disks or was ejected subsequently.  In both cases it is
not very surprising that the specific angular momentum distribution of galactic disks does not agree
with the cosmological predictions which neglect these effects.
\end{abstract}

\begin{figure}[t]
\begin{center}
\includegraphics[width=.9\textwidth]{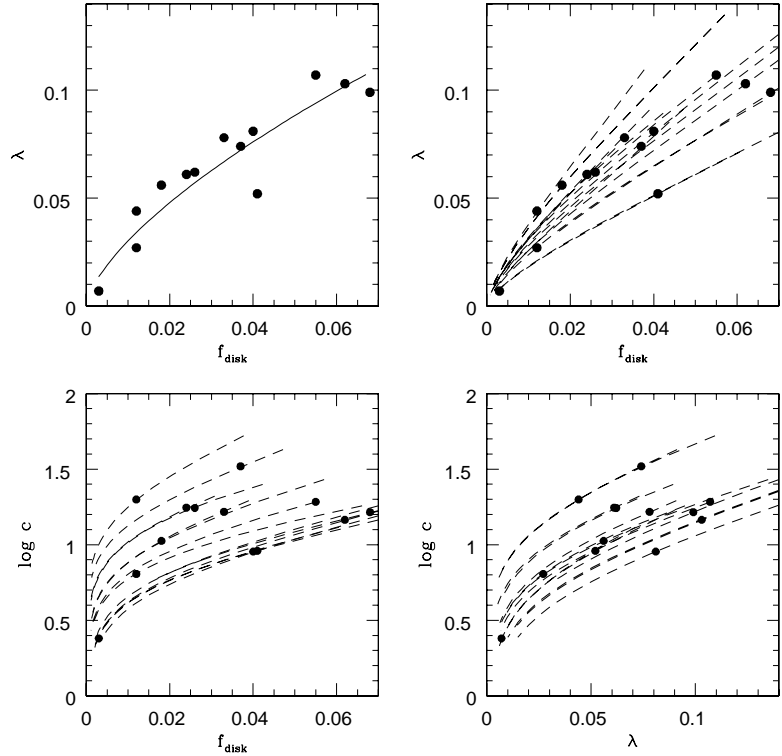}
\end{center}
\caption[]{Correlations between the disk spin parameter, the disk mass fraction and the halo concentration 
for the Swaters galaxy sample. Data points in the upper left panel show the best fit values with no constraints
on the virial parameters, adopting a stellar mass-to-light ratio $\Upsilon_R=1.0(M/L)_{\odot}$. 
The solid line in the upper left panel shows the correlation that would result
from errors in determining the virial radii of dark matter halos. The dashed curves in the other panels
show for each galaxy how the data points shift if one determines the best fitting rotation curve for
different values of the virial radius.}
\label{eps1}
\end{figure}

\section{Introduction}
Within the framework of hierarchical cosmological structure formation galactic disks form
from gas that falls into dark matter halos, where it cools and settles into the equatorial plane.
The disk scale lengths and their rotation curves are determined by the gravitational potential
and by their specific angular momentum distribution which has been acquired from cosmological torques
(Hoyle 1953; Peebles 1969) and the random merging of subunits (Maller, Dekel \& Somerville 2001)
with additional modification during the dissipative protogalactic collapse
phase. Cosmological simulations (Van den Bosch et al. 2002)  have shown that the initial angular momentum distribution 
of the baryonic and dark matter component is similar. This initial condition 
could explain the observed scale lengths and various other properties of galactic disks, provided
that the disk material retained its initial specific angular momentum when settling into the galactic
plane (e.g. Fall \& Efstathiou 1980; Mo, Mao \& White 1998; Firmani \& Avila-Reese 2000; van den Bosch 2001; 
Buchalter, Jimenez \& Kamionkowski 2001). 

In the past couple of years,  high-resolution cosmological NBody/SPH simulations have however uncovered problems with this scenario. Baryons tend to
lose a large fraction of their angular momentum to the dark matter while settling into a disk component. As a result,
simulated galactic disks are an order of magnitude smaller than observed (Navarro \& Benz 1991; Navarro \& Steinmetz 1997). Even if the angular momentum would be conserved, the observed disk angular
momentum distribution does not agree with theoretical predictions. This was shown by
van den Bosch, Burkert \& Swaters (2001), who measured in detail the angular momentum distribution for
a sample of dwarf disk galaxies by fitting a NFW profile (Navarro, Frenk \& White 1997) to the
observed rotation curves, taking into account the disk stars and the HI gas and considering
adiabatic contraction and beam smearing. They confirmed that the mean specific angular momentum of the disk material
is of the same order as expected if angular momentum is conserved during the protogalactic collapse phase.
Their case by case studies however revealed a mismatch of the specific angular momentum profiles of
galactic disks, compared with the predicted universal dark halo angular momentum distribution of Bullock et al. (2001):
the cosmologically predicted mass fraction with low angular momentum is much larger than observed.

The problem of angular momentum loss during gas infall might partly be solved by energetic feedback processes.
Thacker \& Couchman (2001), for example,  showed that stellar heating could decouple the dynamical evolution
of the protogalactic gas with respect to the dark halo,
leading to galactic disks with a specific angular momentum that is within 10\% of the observed value 
(see also Sommer-Larsen et al. 1999, Navarro \& Steinmetz 2000, Maller \& Dekel 2002). Still, as demonstrated by 
van den Bosch et al. (2002), a large fraction
of the baryonic component would have very low or even negative specific angular momentum,
in contrast with the observed disk angular momentum distribution. It has also been suggested that this low-angular momentum gas
could form large galactic bulges instead of disks (Thacker \& Couchman 2002, van den Bosch et al. 2002). These bulges 
are however not observed in the LSB galaxies, studied by Van den Bosch, Burkert \& Swaters (2001).

In addition, van den Bosch, Burkert \& Swaters (2001) detected a strong correlation between the disk spin parameter 
and the disk mass fraction for the Swaters sample. A similar correlation for a much larger sample
of LSB and HSB galaxies has been found by Jimenez, Verde \& Oh (2002). This result is puzzling.
It is not clear why the fraction of baryonic material that forms the observed galactic disks should correlate with the
disk spin parameter.

\section{The origin of the correlation between spin and disk mass fraction}

The upper left panel of figure 1 shows the observed correlation between the disk spin parameter 

\begin{equation}
\lambda = \gamma \frac{j_{tot}}{\sqrt{2}R_{vir}V_{vir}} 
\end{equation}

\noindent and the disk mass fraction $f_{disk} = \frac{M_{disk}}{M_{vir}}$, with
$M_{disk}$ the total disk mass and $M_{vir}=V_{vir}^2R_{vir}/G$ the virial mass of the dark halo.
Here $\gamma$ is a geometrical factor which depends on the dark matter density distribution, 
$j_{tot}$ is the observed total disk angular momentum
and $R_{vir}$ and $V_{vir}$ are the virial radius and virial mass of the dark halo, respectively.
These values represent the best fit to the rotation curves if no constraints are imposed on $R_{vir}$ and
$V_{vir}$.
A dependence of disk rotation on the disk mass fraction might provide interesting new insight 
into the evolution of disk galaxies. However it also could emerge from uncertainties in
determining the dark halo properties.
All the information about the structure of the dark matter halos
is gained through disk rotation curves which are restricted to the inner halo regions. The outer halo regions and
especially their virial masses or virial radii are poorly constrained. In addition, tests show
that the fits to the observed rotation curves are almost equally good for a large range of
virial values.  Both, $\lambda$ and $f_{disk}$ depend on $R_{vir}$.
As $\gamma$ does not vary strongly with halo mass and with $V_{vir} \sim R_{vir}$ we find
$\lambda \sim R_{vir}^{-2}$ and $f_{disk} \sim R_{vir}^{-3}$. 
Any error in $R_{vir}$ will therefore shift the data points along a curve
$\lambda \sim f_{disk}^{2/3}$ which is shown by the solid curve in the upper left panel of figure 1.
The good agreement of the distribution of the data points with this relationship indicates indeed that the correlation
results from errors in determining $R_{vir}$.
This problem is shown in more details and for each galaxy separately
in the right upper panel of figure 1, where the dashed curves show the
best fit values of $\lambda$ and $f_{disk}$ for all galaxies, adopting different values of $R_{vir}$.

\section{Conclusion}

The correlation between 
$\lambda$ and $f_{disk}$ that has been found by van den Bosch et al. (2001) or Jimenez et al. (2002) 
can be explained as a result of uncertainties in determining the dark halo virial radii or masses.  
Cosmological models predict that most of the protogalactic gas with a cosmological baryon
fraction (for LCDM) of $f_{bar} =  \Omega_{bar}/\Omega_0 \approx 0.13$ loses 90\% of its 
angular momentum  and settles into the equatorial plane, leading to typical values of
$\lambda \approx 0.005$ and $f_{disk} \approx 0.13$. 
Even if the virial radius is unknown
and treated as a free parameter, the upper right panel of figure 1 clearly shows
that these values can be ruled out.

Dark matter halos have typical concentrations of order $c \approx 12 - 15$ which should not be affected
strongly by the dynamical evolution of the  baryonic component in dark matter dominated LSB galaxies. 
The lower panels of figure 1 show that $f_{disk} \approx 0.01 - 0.07 < f_{bar}$ for these
halo concentrations.
The galaxies either lost a substantial fraction of their baryons
in a galactic wind or accreted only a small fraction  of
the gas that has been available initially.
In both cases, there exists no reason why the specific angular momentum distribution of the disk component 
should match the dark halo angular momentum distribution as assumed e.g. by Mo, Mao \& White (1998).
The cosmological angular
momentum problem of disk galaxies might therefore be connected directly with the  origin of their
low baryon fractions.

%


\begin{thebibliography}{8.}
\addcontentsline{toc}{section}{References}

\bibitem{j1} A. Buchalter, R. Jimenez \& M. Kamionkowski: MNRAS \textbf{322}, 43 (2001)
\bibitem{j2} J.S. Bullock, A. Dekel, T.S. Kolatt, A.V. Kravtsov, A.A. Klypin,
C. Porciani \& J.R. Primack: ApJ \textbf{555}, 240 (2001)
\bibitem{j3} S.M. Fall \& G. Efstathiou: MNRAS \textbf{193}, 189 (1980)
\bibitem{j4} C. Firmani \& V. Avila-Reese: MNRAS \textbf{315}, 457 (2000)
\bibitem{j5} F. Hoyle: ApJ \textbf{118}, 513 (1953)
\bibitem{j6} R. Jimenez, L. Verde \& S.P. Oh: astro-ph/0201352 (2002)
\bibitem{j7} A.H. Maller, A. Dekel \& R.S. Somerville: MNRAS \textbf{329}, 423 (2001)
\bibitem{j8} A.H. Maller \& A. Dekel: astro-ph/0201187 (2002)
\bibitem{j9} H.J. Mo, S. Mao \& S.D.M. White: MNRAS \textbf{295}, 319 (1998)
\bibitem{j10} J.F. Navarro \& W. Benz: ApJ \textbf{380}, 320 (1991)
\bibitem{j11} J.F. Navarro \& M. Steinmetz: ApJ \textbf{478}, 13 (1997)
\bibitem{j11a} J.F. Navarro, C.S. Frenk \& S.D.M White: ApJ \textbf{490}, 493 (1997)
\bibitem{j12} J.F. Navarro \& M. Steinmetz: ApJ \textbf{538}, 477 (2000)
\bibitem{j13} P.J.E. Peebles: ApJ \textbf{155}, 393 (1969)
\bibitem{j14} J. Sommer-Larsen, S. Gelato \& H. Vedel: ApJ \textbf{519}, 501 (1999)
\bibitem{j15} R.J. Thacker \& H.M.P. Couchman: astro-ph/0106060 (2001)
\bibitem{j16} F.C. van den Bosch, T. Abel, R.A.C. Croft, L. Hernquist \& S.D.M. White: astro-ph/0201095 (2002)
\bibitem{j17} F.C. van den Bosch: MNRAS \textbf{327}, 1334 (2001)
\bibitem{j18} F.C. van den Bosch, A. Burkert \& R.A. Swaters: MNRAS \textbf{326}, 1205 (2001)


\end{thebibliography}
\end{document}